\documentclass[reprint,twocolumn,superscriptaddress,prl,showpacs,amsmath,amssymb,aps]{revtex4-1}

\usepackage{natbib}	
\usepackage{graphicx,color,rotating}
\usepackage[latin1]{inputenc}
\usepackage{textcomp}
\usepackage{dcolumn}
\usepackage{bm}     
\usepackage{upgreek}
\usepackage{subdepth}  			
\usepackage{multirow}
\usepackage[normalem]{ulem}
\usepackage[latin1]{inputenc}

\usepackage{array}
\newcolumntype{L}[1]{>{\raggedright\let\newline\\\arraybackslash\hspace{0pt}}m{#1}}
\newcolumntype{C}[1]{>{\centering\let\newline\\\arraybackslash\hspace{0pt}}m{#1}}
\newcolumntype{R}[1]{>{\raggedleft\let\newline\\\arraybackslash\hspace{0pt}}m{#1}}

\usepackage{hyperref}						
\usepackage{xcolor}						
\hypersetup{							
    colorlinks,
    linkcolor={blue!80!black},
    citecolor={blue!80!black},
    urlcolor={blue!80!black}
}
\newcommand{\SImum}{\textrm{\textmu{}m}}

\newcommand{\SICel}{^\circ\!\textrm{C}}

\newcommand{\eqlab}[1]{\label{eq:#1}}

\newcommand{\mr}[1]{\ensuremath{\mathrm{#1}}}
\newcommand{\myvec}[1]{\bm{#1}}
\newcommand{\vvv}{\myvec{v}}
\newcommand{\uuu}{\myvec{u}}
\newcommand{\fac}{\myvec{f}_\mr{ac}}
\newcommand{\Eac}{E_\mr{ac}}
\newcommand{\pp}{\partial}
\newcommand{\nablabf}{\boldsymbol{\nabla}}

\begin{document}

\title{Fast microscale acoustic streaming driven by a temperature-gradient-induced non-dissipative acoustic body force}

\author{Wei Qiu}
\email{wei.qiu@bme.lth.se}
\affiliation{Department of Biomedical Engineering, Lund University, Ole R\"{o}mers v\"{a}g 3, 22363 Lund, Sweden}

\author{Jonas Helboe Joergensen}
\email{jonashj@fysik.dtu.dk}
\affiliation{Department of Physics, Technical University of Denmark, DTU Physics Building 309, DK-2800 Kongens Lyngby, Denmark}

\author{Enrico Corato}
\affiliation{Department of Biomedical Engineering, Lund University, Ole R\"{o}mers v\"{a}g 3, 22363 Lund, Sweden}

\author{Henrik Bruus}
\email{bruus@fysik.dtu.dk}
\affiliation{Department of Physics, Technical University of Denmark, DTU Physics Building 309, DK-2800 Kongens Lyngby, Denmark}

\author{Per Augustsson}
\email{per.augustsson@bme.lth.se}
\affiliation{Department of Biomedical Engineering, Lund University, Ole R\"{o}mers v\"{a}g 3, 22363 Lund, Sweden}

\date{16 March 2021}

\begin{abstract}
We study acoustic streaming in liquids driven by a non-dissipative acoustic body force created by light-induced temperature gradients. This thermoacoustic streaming produces a velocity amplitude approximately 50 times higher than boundary-driven Rayleigh streaming and 90 times higher than Rayleigh-B\'enard convection at a temperature gradient of 10 K/mm in the channel. Further, Rayleigh streaming is altered by the acoustic body force at a temperature gradient of only 0.5 K/mm. The thermoacoustic streaming allows for modular flow control and enhanced heat transfer at the microscale. Our study provides the groundwork for studying microscale acoustic streaming coupled with temperature fields.
\end{abstract}

\maketitle


Acoustic streaming describes the steady time-averaged fluid motion that takes place when acoustic waves propagate in viscous fluids. The streaming flow is driven by a nonzero divergence in the time-averaged acoustic momentum-flux-density tensor~\cite{Lighthill1978}. Conventionally  in a homogeneous fluid, this nonzero divergence arises from two dissipation mechanisms of acoustic energy. The first case is the so-called boundary-driven  Rayleigh streaming~\cite{LordRayleigh1884}, in which acoustic energy is dissipated in viscous boundary layers where the velocity of the oscillating fluid changes to match the surface velocity of the channel walls~\cite{Nyborg1958, Hamilton2003} or of suspended objects~\cite{Elder1959, Lee1990, Yarin1999, Tho2007}. This gives rise to velocity gradients that drive the flow~\cite{Schlichting1932}, and is typically observed in standing wave fields in systems of a size comparable to the wavelength~\cite{Muller2013}. In the second type of streaming, known as quartz wind or bulk-driven Eckart streaming~\cite{Eckart1948}, high acoustic wave attenuation yields the gradients that drive the streaming flow, which is typically associated with high-frequency traveling waves~\cite{Eckart1948, Nyborg1953a, Riley2001}. Both cases have been extensively studied, and the phenomenon of acoustic streaming continues to attract attention due its importance in processes related to medical ultrasound~\cite{Marmottant2003, vanderSluis2007, Wu2008a, Doinikov2010}, thermoacoustic engines~\cite{Bailliet2001, Hamilton2003a}, acoustic levitation~\cite{Trinh1994, Yarin1999}, and manipulation of particles and cells in microscale acoustofluidics~\cite{Bruus2011c, Barnkob2012a, Hammarstrom2012, Collins2015, Marin2015, Hahn2015a, Guo2016}.

Recently, we discovered that boundary-driven streaming can be significantly suppressed in inhomogeneous media formed by different solute molecules~\cite{Karlsen2018, Qiu2019}. This is attributed to the acoustic body force $\fac$, also originating from the nonzero divergence in the time-averaged momentum-flux-density tensor, due to gradients in density and compressibility in the fluid~\cite{Karlsen2016, Augustsson2016}, which competes with the boundary-layer streaming stress. The streaming rolls are confined to narrow regions near the channel walls, before the inhomogeneity smears out due to diffusion and advection of the solute. This effect enables the acoustic manipulation of submicrometer particles~\cite{Qiu2020} of biological relevance such as bacteria~\cite{Assche2019}, as well as trapping of hot plasma in gasses~\cite{Koulakis2018}.

In this Letter, we investigate microscale acoustic streaming in a liquid, in which the temperature-dependent compressibility and density have been made inhomogeneous by introducing a sustained temperature gradient. We generate this gradient by light irradiation and absorption, and subsequently measure the streaming driven by the temperature-gradient-induced acoustic body force and call it thermoacoustic streaming. Using our newly developed model for thermovisocus acoustofluidics~\cite{Joergensen2020}, the experimental results are validated and the mechanisms responsible for the thermoacoustic streaming are explained.

Our main findings are: (i) The thermoacoustic streaming begins to disturb the boundary-driven Rayleigh streaming with a temperature gradient as small as 0.5~K/mm, resulting in streaming rolls with complex three-dimensional (3D) patterns. (ii) With a temperature gradient of 10~K/mm, the thermoacoustic streaming velocity is about 50 and 90 times higher than that of the boundary-driven Rayleigh streaming and of the Rayleigh-B\'enard convection, respectively. (iii) In contrast to other types of acoustic streaming, the mechanism driving the thermoacoustic streaming is non-dissipative.

The thermoacoustic streaming should be of considerable fundamental interest to a broad community of researchers working in nonlinear acoustics, thermoacoustics, microscale acoustofluidics, as well as heat transfer. For a microsystem, the advective streaming flow is remarkably high compared to the rate of thermal diffusion, and with a Péclet number \textit{Pe} $\approx 1$, heat transfer is strongly enhanced. Further, our findings pave the way for transient or steady control of the streaming through modulations of the temperature or the acoustic field.

\begin{figure}[!t]
\centering
\includegraphics[height=0.80\columnwidth]{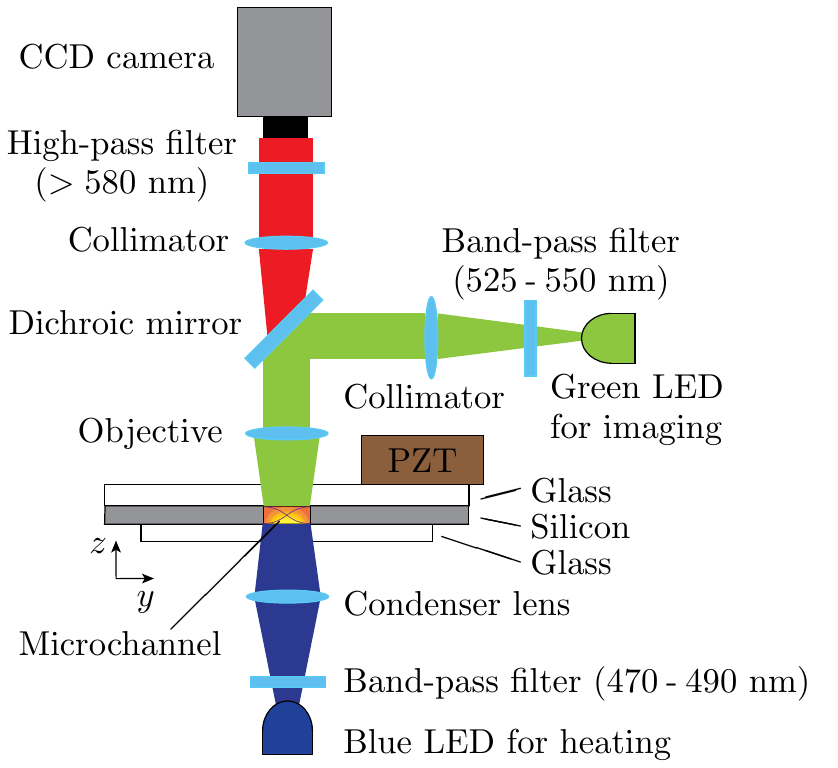}
\caption[]{\label{Sketch}
Sketch of the acoustofluidic silicon chip (light gray) sandwiched between two glass layers (white) that allows optical transparency for both heating and particle tracking. The light emitted from a blue LED below the chip is absorbed by the water containing dye molecules and a temperature gradient forms from low (orange) to high (yellow) temperature in the channel. The piezoelectric transducer (dark brown) excites the resonant half-wave pressure field $p_1$ (purple) at 953~kHz. A green LED shines light from above to excite the red fluorescent light from the tracer particles, which allows the optical recording of the tracer bead motion in a part of length $L = 1300~\SImum$ of the channel of width $W = 760~\SImum$, and height $H = 370~\SImum$.}
\end{figure}

\textit{Experimental method.---} The experiments were performed using a long straight microchannel of width $W = 760 \,\SImum$ and height $H = 370 \,\SImum$ in a glass-silicon-glass chip with a piezoelectric transducer glued underneath. The temperature gradient inside the channel was generated by directing the focused light from a 488-nm light-emitting-diode (LED) through water with a small amount of added dye (0.1 wt\% Orange G) that has an overlapping absorption peak with the LED, see Fig.~\ref{Sketch}, which results in 99\% of the LED light being absorbed in the liquid. The transducer was driven at a frequency of 953~kHz with an input power of 88 mW, which produced a standing half wave across the width with the acoustic energy density $\Eac = 9.24$~J/m$^3$ at room temperature. The induced streaming was measured using general defocusing particle tracking ~\cite{Barnkob2015, Barnkob2020} at 5 to 60~fps with 1.1 $\SImum$-diameter polystyrene tracer particles (red fluorescence). After a band-pass filter (525-550 nm), the excitation light of the tracer particles from a green LED is barely absorbed by the dye molecules and hence does not affect the temperature gradient in the channel. The measurements under each condition were repeated 13 to 27 times and recorded in 7800 to 40500 frames to improve the statistics. The temperature field in the channel mid-height plane was imaged using temperature-sensitive fluorescent dye (Rhodamine B), see the Supplemental Material~\footnote{See Supplemental Material at [url] for details about the temperature measurements, the comparison between measured and simulated temperature fields, the numerical model, as well as the measured and simulated Rayleigh-B\'enard convection.}.

\textit{Numerical model.---} The model is a pressure acoustics model including thermoviscous boundary layers as presented in Ref.~\cite{Joergensen2020}, which enables 3D simulations of thermoviscous acoustofluidic devices.
To simulate the long glass-silicon-glass chip, symmetry planes are exploited to only simulate a quarter of the chip. Furthermore, the perfectly matched layer (PML) technique~\cite{Bermudez2007} is used to avoid simulating the entire length of the chip. The solver consists of three steps: (i) Computing the temperature field $T_0$ induced by the LED with an amplitude set to match the observed temperature gradients, (ii) computing the acoustic displacement in the solid $\uuu_1$ and the pressure $p_1$ in the fluid due to an actuation on the glass, and (iii) computing the resulting acoustic streaming field $\vvv_2$. The heating from the LED is modeled with no absorption in the glass and an absorption parameter in the fluid selected to absorb $99\%$ of the light passing through the chip as measured in the experiment. The model is based on perturbation theory, but the highest streaming velocities recorded in the experiments is found to be at the limit of the validity of the model, because there the thermoacoustic streaming begins to alter the temperature field $T_0$. For more details on the numerical model see the Supplemental Material~\cite{Note1}. Due to the inherent difficulty in measuring energy density $\Eac$ at high streaming velocities when temperature gradients are present, the $\Eac$ used in simulations is obtained by matching the experimental streaming velocity amplitude for each temperature gradient.

\begin{figure*}[!t]
\centering
\includegraphics[width=1.0\textwidth]{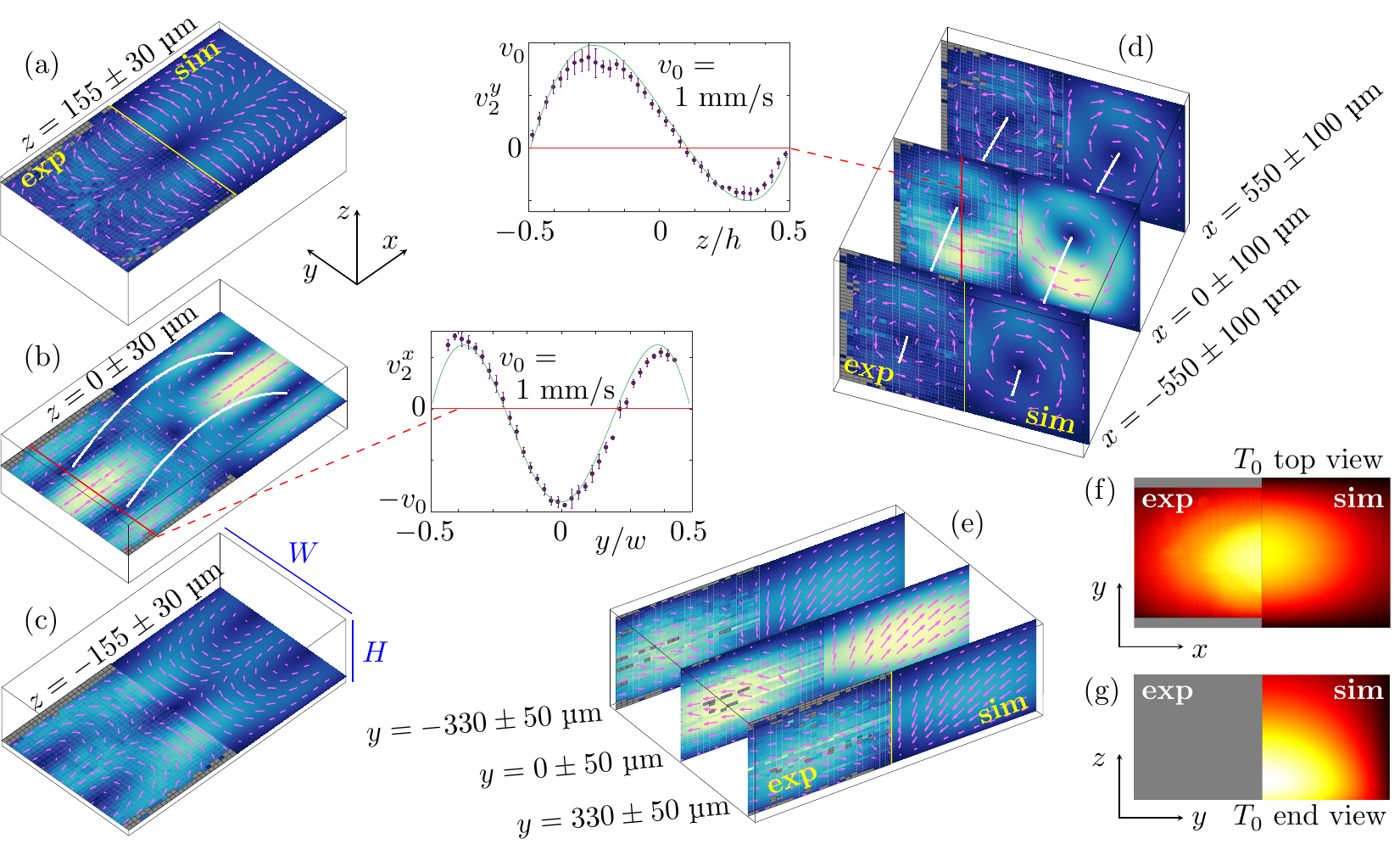}
\caption[]{\label{StreamingPattern}
(a)-(e) The measured (exp, left half) and simulated (sim, right half) streaming patterns for $G =  9.76$~K/mm ($\Delta T_0 = 3.71$~K across the channel width $W$), observed in different planes $x$-$y$ horizontally, $x$-$z$ vertically, and $y$-$z$ vertically. The vector plot (magenta) in a given plane is the in-plane velocity and the color plot is its magnitude from 0 (dark blue) to 1042 $\SImum \, \mathrm{s}^{-1}$ (yellow). The simulation is performed at $\Eac = 23$~J/m$^3$. Spatial bins with no experimental data are excluded (gray). The pair of curved white lines in (b) and (d) represent the centerlines of the two counter-rotating deformed 3D cylindrical streaming flow rolls. The two line-plot insets show the measured (purple) and the simulated (green) $x$ (or $y$) component $v_2^x$ (or $v_2^y$) of the velocity along the red lines. (f) Color plot from 25.0 $\SICel$ (black) to 30.1~$\SICel$ (white) of the measured and simulated temperature $T_0$ in the horizontal $x$-$y$ plane at $z =  0$, see more details in the Supplemental Material~\cite{Note1}. The regions where the fluorescence intensity is affected by the channel sidewalls are excluded (gray). (g) $T_0$ as in panel (f) but for the vertical $y$-$z$ plane at $x = 0$, and only the simulation results are shown.}
\end{figure*}

\textit{Results and discussion.---} When both sound field and temperature gradients are present, the streaming flow exhibits a complex 3D pattern. An example is shown in Fig.~\ref{StreamingPattern}(a)-(e), corresponding to a temperature difference $\Delta T_0$ = 3.71 K across the channel of width $W = 760~\SImum$, equivalent to a gradient $G =2 \Delta T_0/W = 9.76$~K/mm. Here, two counter-rotating deformed 3D cylindrical streaming flow rolls appear, whirling with a velocity amplitude $|{\bm{v}_2}| = 1074~\SImum$/s around the pair of curved white centerlines shown in Fig.~\ref{StreamingPattern}(b) and (d). This velocity amplitude is about 50 and 90 times higher than that of the boundary-driven Rayleigh streaming and the Rayleigh-B\'enard convection, respectively, under the same driving conditions (see Supplemental Material~\cite{Note1}).

The generation of this fast streaming can be explained by the acoustic body force $\fac$ due to the temperature field induced by the blue LED. In this experiment, the light heats the fluid from beneath, while the silicon walls efficiently transport the heat away, thus cooling the sides of the channel. Temperature gradients are therefore induced in all  directions: In the $x$-$y$ plane by the Gaussian profile of the light intensity and by the cooling from the silicon sidewalls, and in the  $z$-direction by light absorption fulfilling the Beer-Lambert law. The resulting temperature field is highest at the center of the channel bottom, as shown by the measured and simulated temperature fields in the horizontal $x$-$y$ plane around channel mid-height $z = 0 \,\SImum$ in Fig.~\ref{StreamingPattern}(f), and by the simulated temperature field in the vertical $y$-$z$ cross section at $x = 0 \,\SImum$ in Fig.~\ref{StreamingPattern}(g). The acoustic body force  $\fac$ depends on the gradients in compressibility and density, the acoustic pressure $p_1$, and the acoustic velocity $\vvv_1$~\cite{Karlsen2016}. When the inhomogeneities are created by a temperature field, $\fac$ can be expressed as,
 \begin{align}
 \fac &=-\frac{1}{4}\vert p_1\vert^2 \nablabf \kappa_{T,0}- \frac{1}{4} \vert \vvv_1\vert^2 \nablabf \rho_0
 \nonumber
 \\
 \eqlab{fac}
 \qquad&
 = -\frac{1}{4} \Bigg[\vert p_1\vert^2 \bigg(\frac{\pp \kappa_T}{\pp T}\bigg)_{T_0} +
 \vert \vvv_1\vert^2  \bigg(\frac{\pp \rho}{\pp T}\bigg)_{T_0}\Bigg]\nablabf T_0.
 \end{align}

Three factors determine the action of $\fac$ on the fluid. (i) Both the compressibility and density decrease with temperature, thus $\fac$ points towards the high temperature region here the center of the channel heated by the LED. (ii) At room temperature, $\kappa_T|p_1|^2 \approx \rho|\bm{v}_1|^2$ and $\frac{1}{\kappa_T}|\pp_T\kappa_T| \gg \frac{1}{\rho}|\pp_T\rho|$, so $\fac$ is dominated by the $\vert p_1\vert^2$ compressibility term and thus is  strongest at the pressure anti-nodes at the sides of the channel. (iii) As shown in  Fig.~\ref{StreamingPattern}(g), the temperature gradient is larger at the bottom than at the top of the channel resulting in a stronger $\fac$ at the bottom. Consequently, in the bottom part of the LED spot, $\fac$ pulls the fluid horizontally inward to the vertical $x$-$z$ center plane at $y = 0$ and by mass conservation lets it escape outward along the axial $x$-direction and upward along the vertical $z$-direction. The resulting streaming flow contains the two aforementioned deformed 3D cylindrical flow rolls, which when projected onto horizontal and vertical planes appear as the four horizontal streaming rolls in Fig.~\ref{StreamingPattern}(a)-(c) strongest in the center-plane $z = 0$, and as the two vertical streaming rolls in Fig.~\ref{StreamingPattern}(d)-(e).

\begin{figure*}[!t]
\centering
\includegraphics[width=1.0\textwidth]{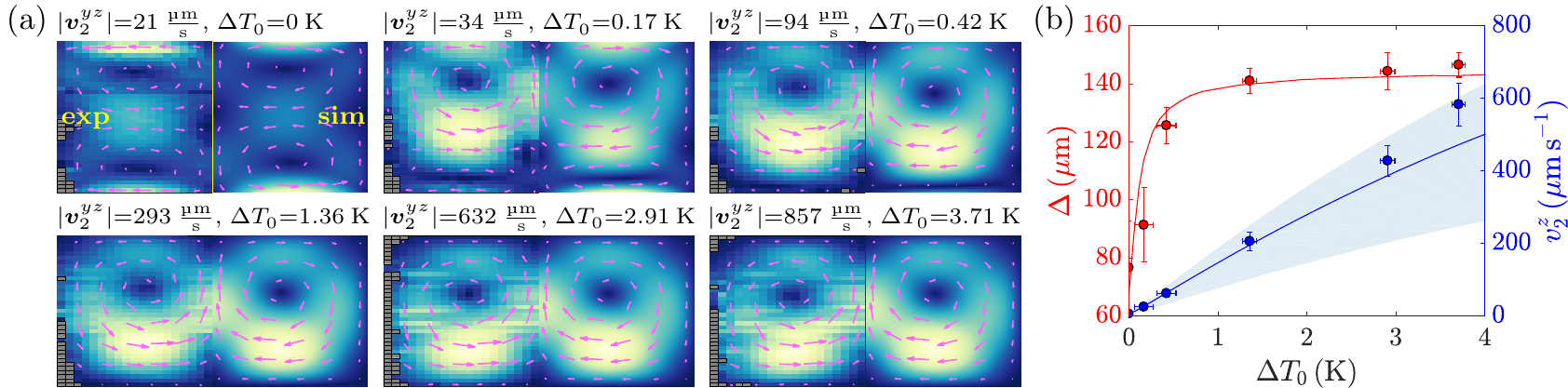}
\caption[]{\label{StreamingEvolution}
(a) Measured (exp, left half) and simulated (sim, right half) in-plane streaming velocity $\bm{v}_2^{yz}$ (magenta vectors) and its magnitude  $|\bm{v}_2^{yz}|$ from 0 (dark blue) to its maximum (yellow) in the vertical $y$-$z$ plane averaged over the vertical slab $x = 0 \pm 100~\SImum$ for six temperature differences across  the channel ranging from $\Delta T_0 = 0$  to 3.71~K. Simulations are performed with the energy density $\Eac$ for which $\left|  {\bm{v}_2^{yz}} \right|$ matches the experimental one under each $\Delta T_0$. Spatial bins with no experimental data points are excluded (gray). (b) Dependence of vortex size $\Delta$ (red) and the velocity in the $z$-direction $\left| {\bm{v}_2^z} \right|$ (blue) on $\Delta T_0$. Simulated $\Delta$ is indicated by a red solid line. Light blue region shows the simulated $\left| {\bm{v}_2^z} \right|$ under $\Eac$ from 9.24 Pa (lower bound) to 23 Pa (upper bound) with a reference (solid blue line) obtained at $\Eac = $ 18 Pa.}
\end{figure*}

In contrast to the inhomogeneity created by diffusing solute molecules out of equilibrium~\cite{Karlsen2018, Qiu2019}, the temperature gradients here are sustained because the system is run in steady state driven by the PZT transducer and the LED light. The resulting streaming is more than one order of magnitude faster than the boundary-driven Rayleigh streaming, and it is mainly due to the large non-dissipative $\fac$. The thermoacoustic driving mechanism is fundamentally different from the dissipation mechanism of the conventional forms of acoustic streaming.

The transition from boundary-driven Rayleigh streaming to thermoacoustic streaming is studied in the vertical $y$-$z$ cross section ($-100 \,\SImum < x < 100 \,\SImum$) by gradually increasing the output power of the LED. Following Refs.~\cite{Karlsen2018, Qiu2019}, we quantify the streaming evolution by the vortex size $\Delta$, defined as the average of the distance from the center of each of the two upper flow rolls to the channel ceiling at $z=\frac12 H$. Figure~\ref{StreamingEvolution} shows that at a zero temperature gradient $G$, the streaming is governed by the four conventional boundary-driven Rayleigh streaming rolls, whereas at high gradient $G \approx 3.6$~K/mm ($\Delta T_0 = 1.36$~K), only two big temperature-gradient induced flow rolls driven by the relatively large $\fac$ is present, occupying the whole channel cross section. In transitioning from the former to the latter situation, the two top Rayleigh flow rolls appear to expand downwards squeezing the bottom rolls against the channel floor at $z = - \frac12 H$. This phenomenon can be explained by the fact that the two top (bottom) Rayleigh flow rolls have the same (opposite) rotation direction as the two temperature-gradient induced flow rolls. Already at $G = 0.5$~K/mm ($\Delta T_0 = 0.170$~K), $\fac$ is large enough to distort the four-flow-roll Rayleigh streaming pattern. When $\Delta T_0$ further increases, the two-flow-roll thermo\-acoustic streaming pattern dominates, and eventually remains unchanged, while the velocity amplitude increases almost linearly for the investigated range, see Fig.~\ref{StreamingEvolution}(b).

The observation that thermoacoustic streaming occurs already at temperature gradients below 0.5 K/mm calls for caution when combining acoustofluidic devices with optical systems. For an absorbing liquid, the light in a standard microscope is enough to induce strong velocity fields that can interfere with the study object. While we did not record the transient build-up of the streaming field upon activating the light, it can be noted that the development of the temperature field, and thus the streaming field, occurs within a few hundred milliseconds at the studied length scale which opens for rapid modulation of local streaming fields through fast re-configurable optical fields. Even so, the induced streaming velocity is high enough to match the thermal diffusion time and thereby impact the heat transfer in the structure.

\textit{Conclusion.---}This letter describes a comprehensive experimental and numerical study of the thermoacoustic streaming in liquids induced by temperature gradients generated by light absorption in a microchannel. We have obtained a good match between measured and simulated velocity fields in 3D. As summarized by the main findings (i)-(iii) in the introduction, the thermoacoustic streaming, driven by the non-dissipative acoustic body force, has a markedly different origin than that of the conventional acoustic streaming associated with energy dissipation. Moreover, it reaches much higher velocity amplitudes compared to the boundary-driven Rayleigh streaming and the Rayleigh-B\'enard convection under comparable conditions. The acoustic body force relies on the acoustic field and the gradients in compressibility and density, which is analogous to the driving force of the classical Rayleigh-B\'enard convection relying on  the gravitational field and the gradient in density.

By including the temperature dependence of both density and compressibility in this work through our theory in Ref.~\cite{Joergensen2020}, our analysis of thermoacoustic streaming in terms of the acoustic body force is valid for both liquids and gases. Thus, we have extended previous related work on gases, where compressibility effects are unimportant and therefore neglected~\cite{Hamilton2003a, Aktas2010, Chini2014, Vcervenka2017, Michel2019, Fand1960, Thompson2005, Thompson2005a, Nabavi2008}.

This study is fundamental in scope, but also demonstrates a method with clear potential for controlling local flows in microchannels. Further, we highlight important implications of this phenomenon relating to heat transfer and integration of optical fields with microscale acoustofluidic devices.

We are grateful to R. Barnkob (Technical University of Munich) and M. Rossi (Technical University of Denmark) for providing the software GDPTlab. W.Q. was supported by the Foreign Postdoctoral Fellowship from Wenner-Gren Foundations and by MSCA EF Seal of Excellence IF-2018 from Vinnova, Sweden's Innovation Agency (Grant No. 2019-04856).
This work has received funding from Independent Research Fund Denmark, Natural Sciences (Grant No. 8021-00310B) and the European Research Council (ERC) under the European Union's Horizon 2020 research and innovation programme (Grant Agreement No. 852590).

%
%


%

\end{document}